\documentstyle[prl,preprint,aps]{revtex}

\newlength{\headroom}
\newlength{\psfigskip}
\begin{document}
\title{
		     Big Bang Nucleosynthesis in Crisis?
}
\author{             N.\ Hata, R.\ J.\ Scherrer, G.\ Steigman,
		     D.\ Thomas, and T.\ P.\ Walker\cite{OJI} }
\address{
{\it
		     Department of Physics,
		     The Ohio State University,               \\
	             Columbus, Ohio 43210                     \\
}}
\author{             S.\ Bludman and P.\ Langacker        }
\address{
{\it
                     Department of Physics,
                     University of Pennsylvania,               \\
                     Philadelphia, Pennsylvania 19104          \\
}}

                     \date{May 16, 1995, Revised: August 24, 1995,
			   OSU-TA-6/95, UPR-0654T, hep-ph/9505319}

\maketitle
%

\renewcommand{\baselinestretch}{1.3}
\begin{abstract}

A new evaluation of the constraint on the number of light neutrino
species (N$_\nu$) from big bang nucleosynthesis suggests a discrepancy
between the predicted light element abundances and those inferred from
observations, unless the inferred primordial $^4$He abundance has been
underestimated by 0.014$\pm$ 0.004 ($1\sigma$) or less than 10\%
(95\%C.L.) of $^3$He survives stellar processing.  With the quoted
systematic errors in the observed abundances and a conservative
chemical evolution parameterization, the best fit to the combined data
is N$_\nu = 2.1 \pm 0.3$ (1$\sigma$) and the upper limit is N$_\nu <
2.6$ (95\% C.L.).  The data are inconsistent with the Standard Model
(N$_\nu = 3$) at the 98.6\% C.L.

\end{abstract}
\pacs{PACS numbers: }
%


Along with the Hubble expansion and the cosmic microwave background
radiation, big bang nucleosynthesis (BBN) provides one of the key
quantitative tests of the standard big bang cosmology.  The predicted
primordial abundances of $^4$He, D, $^3$He, and $^7$Li
\cite{WSSOK,Smith-Kawano-Malaney} have been used to constrain the
effective number of light neutrino species (N$_\nu$)
\footnote{ %
Neglecting the baryon contribution, the total energy density
$\rho_{\rm tot}$ depends on N$_\nu$ as $\rho_{\rm tot} =
\rho_\gamma + \rho_e + {\rm N}_\nu \rho_\nu$, where $\rho_\gamma$,
$\rho_e$, and $\rho_\nu$ are the energy density of photons, electrons
and positrons, and massless neutrinos (one species), respectively. }
\cite{Steigman-Schramm-Gunn,1989,WSSOK,Kernan-Krauss}.  The neutrino
counting includes anything beyond the Standard Model [such as
a right-handed (sterile) neutrino] that contributes to the energy
density.  This constraint is complementary to neutrino counting
from the invisible width of $Z$ decays (N$_\nu^Z$), which is sensitive
to a much larger mass range ($\lesssim M_Z/2$, where $M_Z$ is the $Z$
mass), but only to neutrinos fully coupled to the $Z$; the current
result is N$_\nu^Z = 2.988 \pm 0.023$ \cite{PDG}, in agreement with
the Standard Model (N$_\nu^Z$ = 3).

The primordial $^4$He abundance is sensitive to the competition
between the early universe expansion rate and the weak interaction
rates responsible for the interconversion of neutrons and protons.
The expansion rate depends on the overall density and hence on
N$_\nu$, while the weak rates are normalized via the neutron lifetime.
Recent improvements in neutron lifetime measurements have
significantly reduced the uncertainty in the $^4$He prediction and,
coupled with increasingly accurate astronomical data on extragalactic
$^4$He, have led to tighter constraints on N$_\nu$; at 95\% C.L.\
N$_\nu < 4$ in 1989 \cite{1989}, $< 3.3$ in 1991 \cite{WSSOK}, and $<
3.04$ in 1994 \cite{Kernan-Krauss}.  However, a constraint as strong
as N$_\nu < 3.04$ hints that the standard theory with N$_\nu = 3$ may
not provide a good fit to the observations.

In this {\it Letter} we present new BBN limits on N$_\nu$ and the
baryon-to-photon ratio ($\eta$) from simultaneous fits to the
primordial $^4$He, D, $^3$He and $^7$Li abundances [hereafter we use
the notation Y$_{\rm p}$ ($^4$He mass fraction), $y_{2{\rm p}}$ = D/H,
$y_{3{\rm p}}$ = $^3$He/H, and $y_{7{\rm p}}$ = $^7$Li/H, fractions by
number] inferred from the astrophysical observations.  In particular,
we incorporate new constraints on $y_{2{\rm p}}$ \cite{D-paper}, which
are based on a generic chemical evolution parameterization \cite{ST95}
and which significantly improve the prior constraints
\cite{YTSSO,WSSOK}.  Our likelihood analysis systematically
incorporates the theoretical and observational uncertainties.  The
theoretical uncertainties and their correlations are estimated by the
Monte Carlo method
\cite{Krauss-Romanelli,Kernan-Krauss,Copi-Schramm-Turner,SBBN-analysis}.
Non-Gaussian uncertainties in the observations, such as the adopted
systematic error in the value of Y$_{\rm p}$, the upper and lower
limits for D, and the model-dependent $^3$He survival parameter
($g_3$), are treated in a statistically well-defined way.

We adopt a primordial helium abundance estimated from low metallicity
HII regions \cite{Olive-Steigman-He4}:
\begin{equation}
        {\rm Y}_{\rm p} = 0.232 \pm 0.003\, ({\rm stat})
                                \pm 0.005 \, ({\rm syst}),
\end{equation}
assuming a Gaussian distribution for the 1$\sigma$ statistical
uncertainty and a flat (top hat) distribution with a half width of
0.005 for the systematic uncertainty \cite{SBBN-analysis}.  The
systematic error is similar to that used for previous estimates on
N$_\nu$ \cite{1989,WSSOK,Kernan-Krauss} and to that obtained from
Pagel's analysis of the data \cite{Pagel}.

New D constraints were obtained in Refs.~\cite{D-paper,ST95}, using
pre-solar abundances of D and $^3$He (as inferred from $^3$He
measurements in the solar wind, meteorites, and lunar
soil\cite{Geiss}) and a generic chemical evolution parameterization:
\begin{eqnarray}
        y_{2{\rm p}} &=& (1.5 - 10.0) \times 10^{-5}  \\
        y_{3{\rm p}} &\le& 2.6 \times 10^{-5} \hspace{3em}(95\%\ \mbox{C.L.}).
\end{eqnarray}
Although these constraints are independent of any specific model for
primordial nucleosynthesis, standard BBN or otherwise, they do depend
on the adopted $^3$He survival fraction $g_3$.  To be consistent with
prior analyses we adopt $g_3 = 0.25$ \cite{YTSSO,DSS,WSSOK} although
the effective $g_3$ of most models is significantly larger than this
(see later discussion).  When the observational bounds in Eqs.~2 and 3
are convolved with the BBN predictions (which are a function of $\eta$
with N$_\nu$ fixed at 3), even tighter constraints on D and $^3$He may
be inferred \cite{D-paper}: $y_{2{\rm p}} = (3.5^{+2.7}_{-1.8}) \times
10^{-5}$ and $y_{3{\rm p}} = (1.2 \pm 0.3) \times 10^{-5}$ at 95\%
C.L..  The resulting upper bound to $y_{2{\rm p}}$ is roughly 30\%
lower than the corresponding bound in Ref.~\cite{WSSOK} and this has
the effect of raising the lower bound on the allowed range of $\eta$.
Our central value for $y_{2{\rm p}}$ is an order of magnitude smaller
than the abundance inferred from a possible D detection in absorption
against a high redshift QSO \cite{Songaila-etal,Carswell-etal}, but
consistent with that reported for a different QSO absorption system
\cite{Tytler}.

We estimate the primordial $^7$Li abundance from the metal-poor stars
in our Galaxy's halo:
\begin{equation}
        y_{7{\rm p}} = (1.2 ^{+4.0}_{-0.5}) \times 10^{-10}
                        \hspace{3em}(95\%\ \mbox{C.L.}).
\end{equation}
This estimate is consistent with other recent determinations
\cite{Li,Copi-Schramm-Turner} which take into account possible post
big bang production and stellar depletion of $^7$Li.

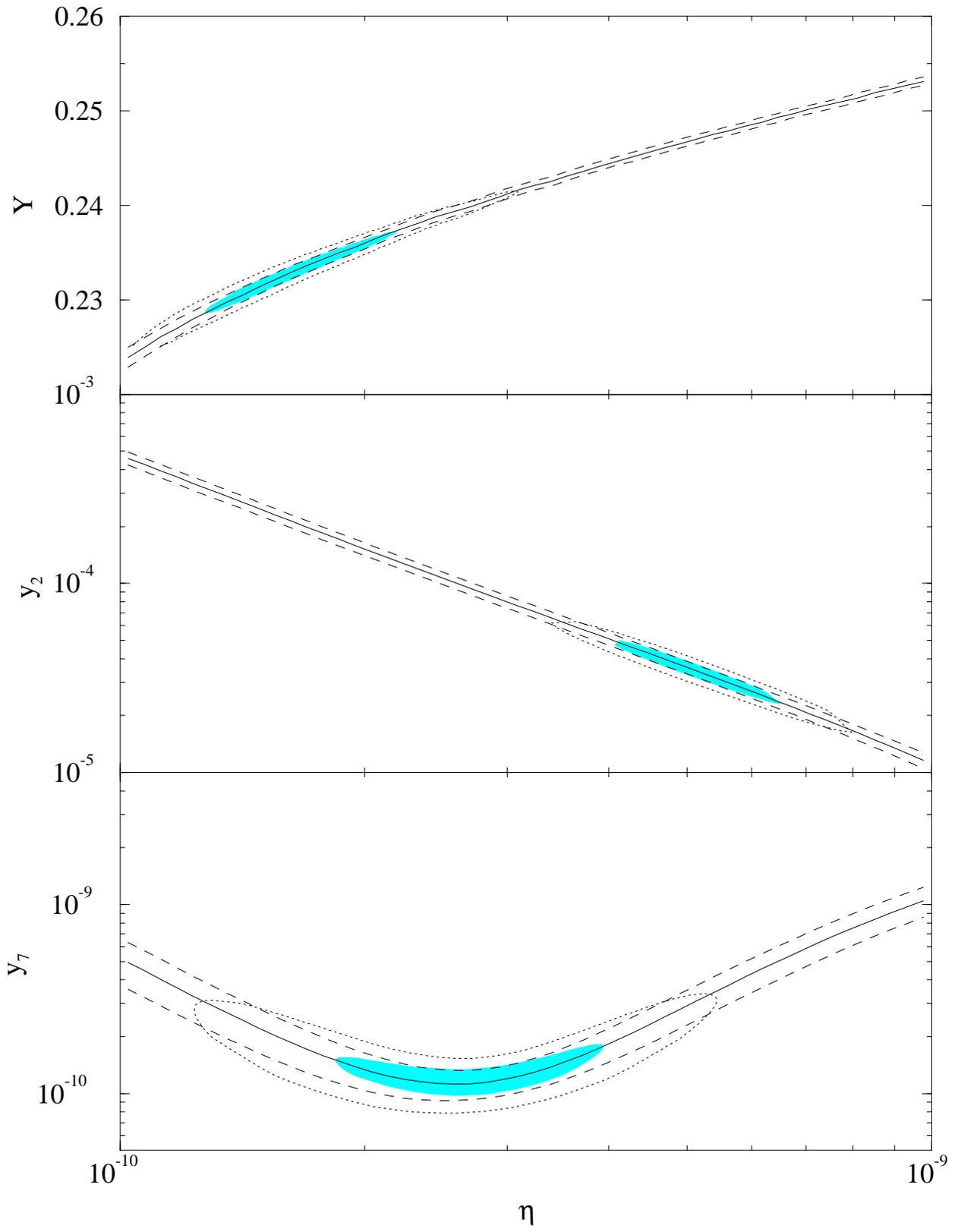
\begin{figure}[h]

\caption{
BBN predictions (solid lines) for Y$_{\rm p}$, $y_{2{\rm p}}$, and
$ y_{7{\rm p}}$ with the theoretical
uncertainties (1$\sigma$) estimated by the Monte Carlo method (dashed
lines).  Also shown are the regions constrained by the observations at
68\% and 95\% C.L. (shaded regions and dotted lines, respectively).}
\label{fig:bbn_comb}
\end{figure}

For standard (N$_\nu = 3$) BBN, the theoretical predictions with the
uncertainties (1$\sigma$) determined by the Monte Carlo technique are
displayed as a function of $\eta$ in Fig.~\ref{fig:bbn_comb}.  Also
shown in Fig.~\ref{fig:bbn_comb} are the constraints obtained by our
likelihood analysis of the predictions and observations.  The result
is disturbing: the constraints on $\eta$ from the observed $^4$He and
D$-$$^3$He abundances appear to be mutually inconsistent.

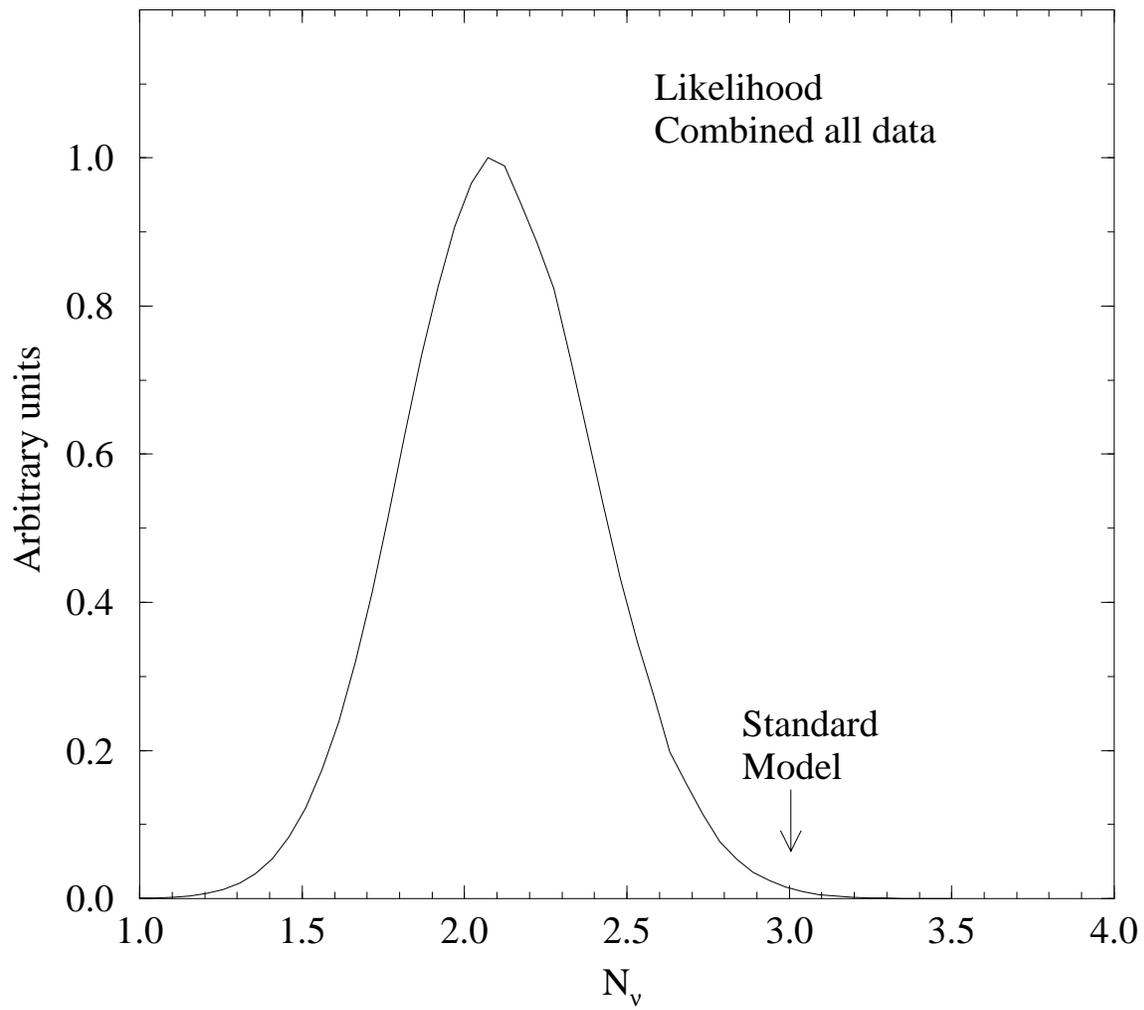
\begin{figure}[h]

\caption{
The likelihood function for N$_\nu$ when the observations
for Y$_{\rm p}$, $y_{2{\rm p}}$, $y_{3{\rm p}}$, and
$ y_{7{\rm p}}$ are fit
simultaneously.  For each N$_\nu$ the likelihood function is maximized
for $\eta$.  The upper limit is N$_\nu < 2.6$ (95\% C.L.)\,
The fit for the Standard Model (N$_\nu = 3$) is excluded at 98.6\% C.L.}
\label{fig:nnu_P}
\end{figure}

To explore this more carefully, all four elements are fit
simultaneously, yielding the likelihood function for N$_\nu$ shown in
Fig.~\ref{fig:nnu_P} (where the likelihood is maximized with respect
to $\eta$ for each N$_\nu$).  The BBN predictions for the D, $^3$He,
and $^7$Li abundances are sensitive to the baryon-to-photon ratio
$\eta$, but only weakly dependent on N$_\nu$.  The BBN prediction for
$^4$He is very weakly dependent on $\eta$ and is approximately
proportional to (${\rm N}_\nu-3$).  In our likelihood analysis, we have
computed the Monte Carlo predictions for all of the element abundances
for $1.5 \le {\rm N}_\nu \le 4$ and $10^{-10} \le \eta \le 10^{-9}$.
The N$_\nu$ and $\eta$ dependences of the uncertainties, the $\eta$
dependence of the correlations among the uncertainties
\cite{Walker,Kernan-Krauss,SBBN-analysis}, and the correlations between
$\eta$ and the $y_{2{\rm p}}$ and $y_{3{\rm p}}$ values have all been
included in the likelihood function.

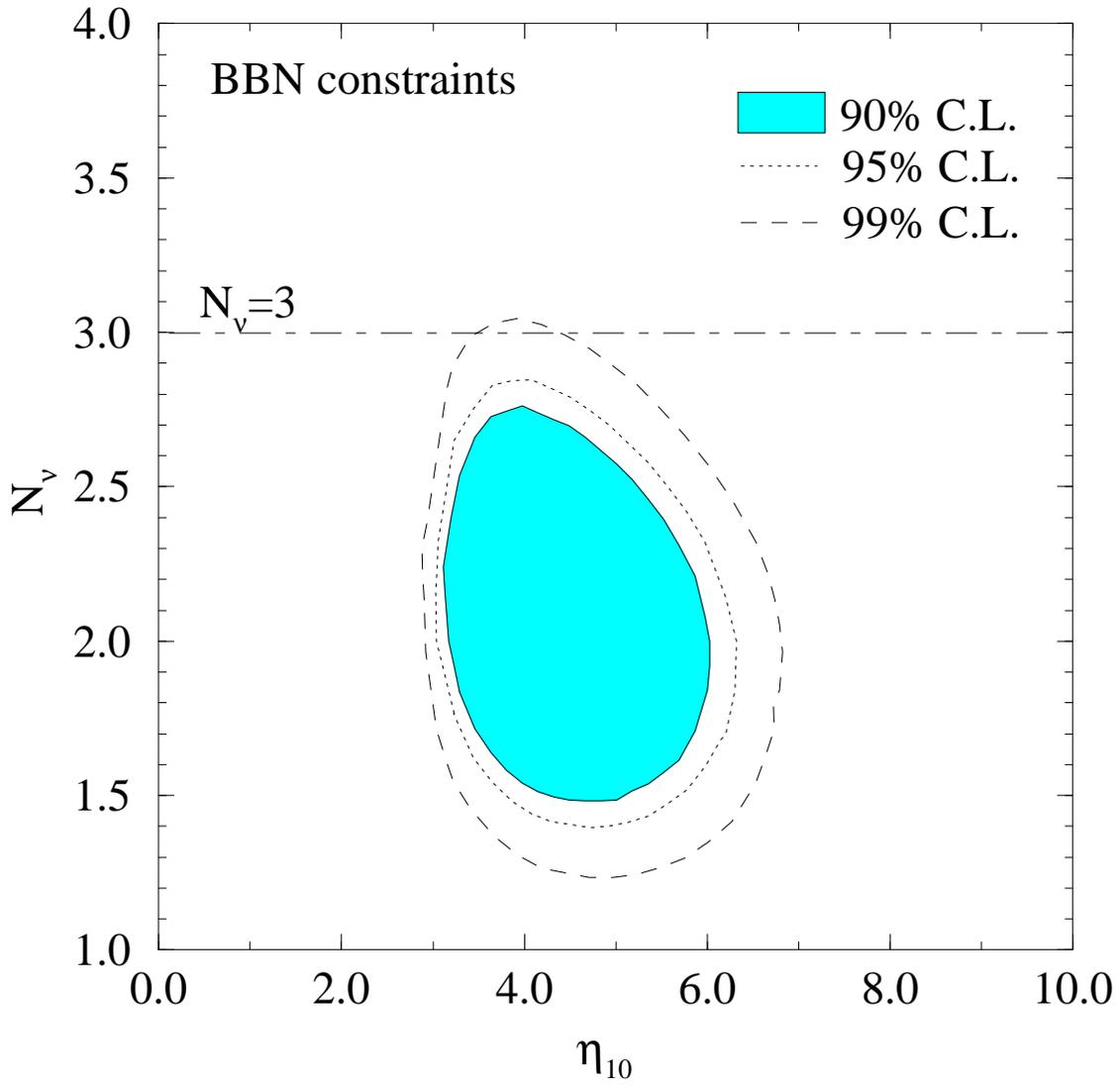
\begin{figure}[h]

\caption{
The combined fit of the observations to N$_\nu$ and $\eta_{10}
\equiv \eta\; 10^{10}$.  }
\label{fig:nnu}
\end{figure}

Fig.~\ref{fig:nnu_P} shows that the Standard Model (N$_\nu = 3$)
yields an extremely poor fit.  The best fit is for N$_\nu = 2.1 \pm
0.3$, and the upper-limit from the joint likelihood
(Fig.~\ref{fig:nnu_P}) is
\begin{equation}
        {\rm N}_\nu < 2.6  \mbox{ (95\% C.L.)}.
\end{equation}
The ratio of the likelihood of N$_\nu = 3$ to the best fit N$_\nu$ =
2.1 is 0.014.  This value provides an estimate of the
goodness-of-fit of the standard (N$_\nu=3$) theory.
\footnote{%
There is no standard procedure to estimate the goodness-of-fit when
non-Gaussian uncertainties are involved in a likelihood analysis.  In
addition to using the ratio of the likelihoods for N$_\nu =2.1$ and 3,
we have also estimated the goodness-of-fit with the standard $\chi^2$
method by approximating the errors with Gaussian distributions: the
results from the two methods are consistent \cite{SBBN-analysis}.  }
The result of our simultaneous fit in the $\eta -{\rm N}_\nu$ plane is
shown in Fig.~\ref{fig:nnu}.  The constraint on the baryon-photon
ratio is $\eta = (4.4^{+0.8}_{-0.6}) \times 10^{-10}$ (1$\sigma$).
The conflict between the lower and upper bounds on $\eta$ coming from
D and $^4$He, respectivley, has been noted before \cite{BBN-problem}.
Our results exacerbate this discrepancy to roughly a 3 standard
deviation effect, mainly due to our new D constraint.

In setting limits when the likelihood function extends beyond the
physical parameter space, it is usually a reasonable (and
conservative) prescription to renormalize the probability density
distribution within the physical part of parameter space.  This
implies that one should renormalize the likelihood function for N$_\nu
\ge 3$, when constraining any (nonstandard) particle contribution in
addition to three massless neutrinos in the Standard Model.  Examining
the N$_\nu$ limit this way, the 95\% C.L. limit for N$_\nu$ extends to
3.25 (for $\eta = 4.6 \times 10^{-10}$).  However, we do not advocate
this interpretation, since the poorness of the N$_\nu = 3$ fit makes
this additional constraint for N$_\nu > 3$ meaningless.

The combined data (D, $^3$He, $^4$He, and $^7$Li) with the adopted
uncertainties are inconsistent with standard (N$_\nu = 3$) BBN, for a
conservative choice of $^3$He survival factor $g_3=0.25$.  But what if
some of the uncertainties have been underestimated?  In particular,
the systematic uncertainty in the $^4$He observational data may be 3
or more times larger \cite{Sasselov-Goldwirth} than the estimate in
Ref.~\cite{Olive-Steigman-He4}.  With $\eta$ determined by the
combined D$-$$^3$He and $^7$Li constraints, BBN predicts $Y_p= 0.246
\pm$ 0.002 (1$\sigma$), where the error includes the uncertainties
from the D$-$$^3$He and $^7$Li constraints and from the BBN theory
calculation.  This value for $Y_p$ required for BBN consistency is
0.014 above the adopted observed value [Eq. (1)].

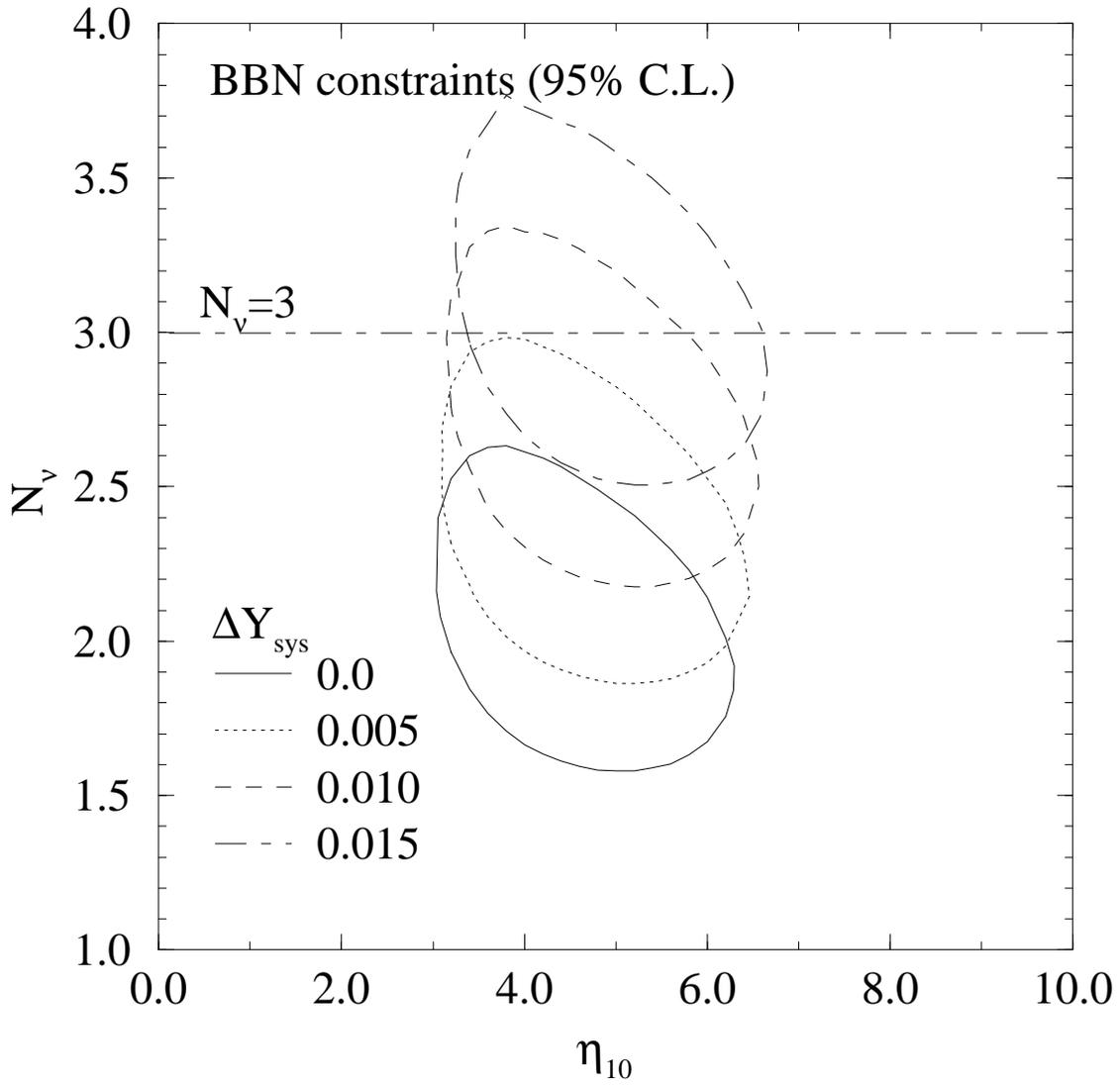
\begin{figure}[h]

\caption{
The combined fit of the observations when the systematic uncertainty in
the $^4$He observation ($\Delta {\rm Y}_{\rm sys}$) is fixed to 0,
0.005, 0.010, and 0.015.   }
\label{fig:nnu_dYsys}
\end{figure}

In Fig.~\ref{fig:nnu_dYsys} we show the $\eta-{\rm N}_\nu$ constraints
when the central value for $Y_p$ is systematically shifted by $\Delta
{\rm Y}$.  To be consistent with N$_\nu = 3$, $\Delta {\rm Y}$ has to
be significantly larger than the systematic error adopted in Eq.~1.
When $\Delta {\rm Y}$ is fit as a free parameter with N$_\nu$ fixed to
3, we obtain $\Delta {\rm Y} = 0.014 \pm 0.004$ at 1$\sigma$.  Even
allowing $\Delta {\rm Y}$ to change freely, the $^7$Li and ISM D
constraints still bound $\eta$ from above at $6.3 \times 10^{-10}$
(95\% C.L.); ISM D alone bounds $\eta$ from above at $9 \times
10^{-10}$.  The claim in Ref.~\cite{Sasselov-Goldwirth} that $\eta$
can be as large as $\sim 14 \times 10^{-10}$ is unjustified.

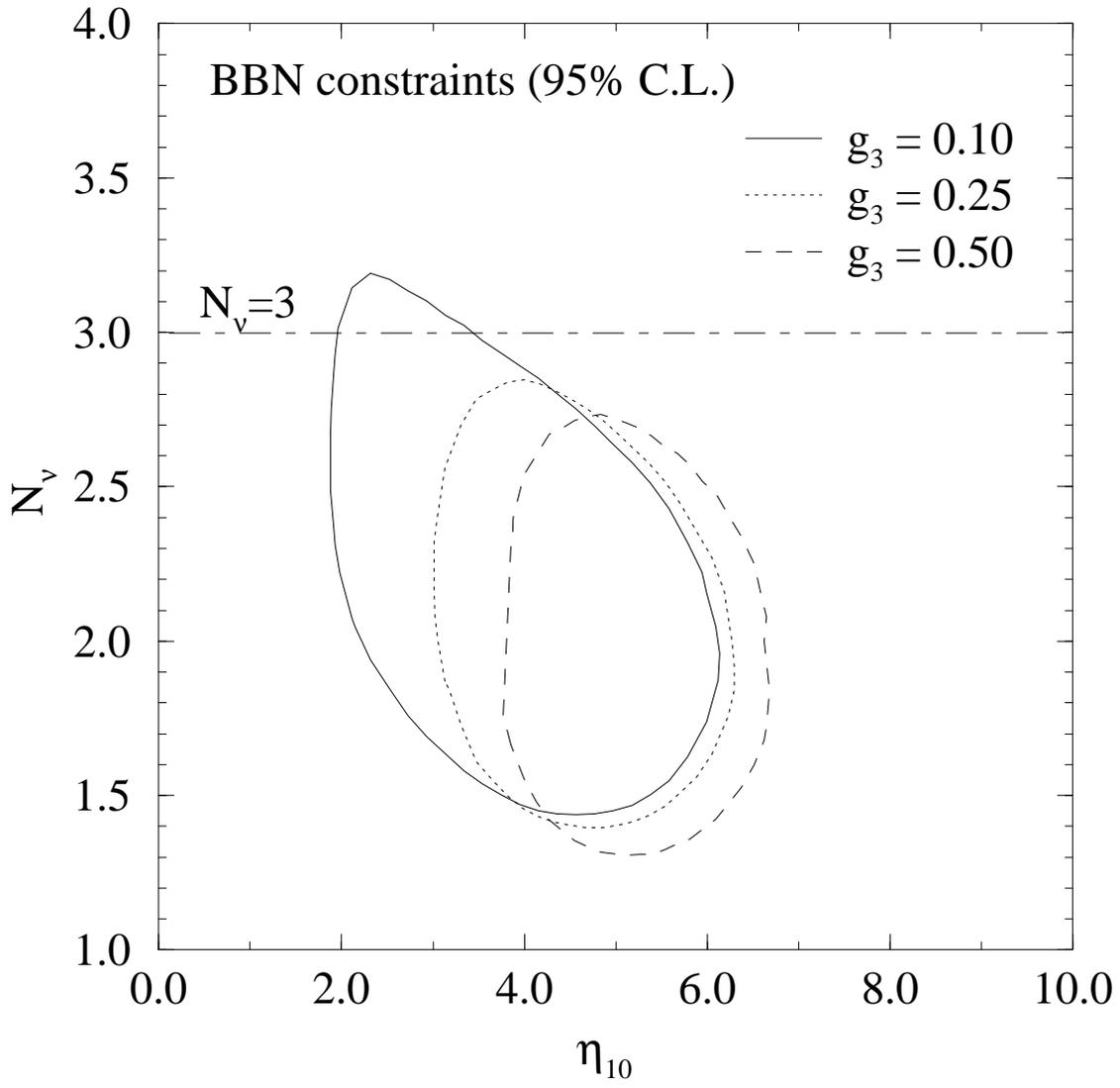
\begin{figure}[h]

\caption{
The combined fit of the observations when the $^3$He survival factor
($g_3$) is fixed to 0.10, 0.25, and 0.50.}
\label{fig:nnu_g3s}
\end{figure}

We have also examined (Fig.~\ref{fig:nnu_g3s}) how the $\eta-{\rm
N}_\nu$ constraint is relaxed when the $^3$He survival factor, which
affects the upper limit on $y_{2{\rm p}}$, differs from that adopted
($g_3 = 0.25 $).  Relaxing the $y_{2{\rm p}}$ upper limit so as to be
consistent with the Y constraint requires a significantly smaller
$g_3$.  When $g_3$ is allowed to be a free parameter with N$_\nu$= 3
fixed, we obtain $g_3 \le 0.10$ at 95\% C.L., i.e. stellar destruction
of $^3$He would need to be significantly larger than is implied by
stellar and chemical evolution models.  Although it is difficult to
assign statistical probabilities to various values of $g_3$, one can
assess the current status of models of Galactic chemical evolution and
their associated $^3$He destruction.  In this {\sl Letter} we have
adopted an effective\footnote{The $g_3$ used in previous BBN analyses
is an {\it effective} $g_3$ in that it represents the $g_3$ per star
integrated over all stars and cycled thru some number of stellar
generations.} $g_3= 0.25$, a choice based on the fact that $g_3\ge
0.25$ for {\sl any} star \cite{DSS,YTSSO,WSSOK}.  Recent studies
\cite{ST92,VOP,ST95} have effective $g_3$'s larger than 0.25, a fact
supported by Ostriker and Schramm's analysis of horizontal branch
stars\cite{OS} which concludes that $g_3>0.3$ and Rood, Bania and
Wilson's observation of $^3$He in planetary nebulae which suggests
that low mass stars are net producers of $^3$He\cite{RBW}.  In order
for the effective $g_3$ to be lower than 0.25, gas would have to be
cycled thru several generations of relatively massive stars (which are
the most efficient destroyers of $^3$He) without overproducing
metals. Allowing stellar $^3$He production (as evidenced in low mass
stars) would effectively increase $g_3$ and therefore exacerbate the
present discrepancy between theory and observations.  There are models
and parameterizations which attempt to address these issues.  The
models of Olive {\it et al.} \cite{ORSTV} include stellar $^3$He
production in low mass stars and therefore tend towards large $g_3$.
They conclude that ``the only way to reduce $g_3$ below that of the
massive stars (around 0.3) would be to argue that the gas in the
region has been cycled through stars several times.  Such an
assumption however would invariably predict $^4$He abundance factors
of 2-4 higher than those observed.'' Vangioni-Flam and Casse
\cite{VFC} find that the effective $g_3$ can be small, but the
associated metals overproduction requires the revision of classical
models of chemical evolution ({\it e.g.}, including metal depletion by
outflow).  The interplay between the lower bound to $g_3$ and metal
overproduction is reflected in Copi, Schramm and Turner's
\cite{CST-II} `stochastic history' parameterization of chemical
evolution.  Their 95\% C.L. lower bounds to $\eta$ are greater than or
equal to ours provided they satisfy the metallicity constraint.  It is
our conclusion that our D constraint is robust and probably overly
conservative - most models of chemical evolution yield D constraints
which make the fit between theory and observation for N$_\nu$ = 3
worse than we report here.  For example, if we assume that $g_3$ is
equally likely to be between 0.25 and 0.5, standard BBN would be ruled
out at the 99.1\% C.L..

The standard (N$_\nu = 3$) BBN predictions for the primordial $^4$He
and D abundances appear to be inconsistent with those inferred from
observations, unless the inferred primordial $^4$He mass fraction has
been underestimated by $\Delta {\rm Y} = 0.014 \pm 0.004$ or the
$^3$He survival fraction, $g_3$, is smaller than 0.10.  While it may
be that the crisis lies in the observational data and/or its
extrapolation to primordial abundances, it is possible to alter
standard BBN in order to reduce the $^4$He prediction to the level
consistent with the D constraint.  The effective N$_\nu$ can be
reduced to the range $2.1 \pm 0.3$ in several ways: massive tau
neutrinos, neutrino degeneracy, or new decaying particles to name but
a few.


This work is supported by the Department of Energy Contract No.\
DE-AC02-76-ER01545 at Ohio State University and DE-AC02-76-ERO-3071 at
The University of Pennsylvania.  R.J.S. is supported in part by NASA
(NAG 5-2864).  We thank C. Copi, D. Schramm, and M. Turner for
spirited discussions relating to this paper and others.

%
%



%
%

%
%

%
%

%
%

%
%

\end{document}